\documentclass[11pt]{article}
\usepackage[preprint]{acl}
\usepackage{times}
\usepackage{latexsym}
\usepackage[T1]{fontenc}
\usepackage[utf8]{inputenc}
\usepackage{microtype}
\usepackage{graphicx}
\usepackage{multirow}
\usepackage{booktabs}
\usepackage{amsmath}
\usepackage{tikz}
\usepackage{subcaption}
\usetikzlibrary{arrows.meta,positioning}
\usepackage{enumitem}
\usepackage{soul, xcolor}
\usepackage{tcolorbox}
\usepackage{colortbl}
\usepackage{amssymb}

\title{Why Can't They Remember? Uncovering Representation and Retrieval Bottlenecks in Multi-Turn Acoustic Memory}

\author{
  Yang Xiao$^1$, Siyi Wang$^1$, Han Yin$^2$, Hong Jia$^3$, 
  Vidhyasaharan Sethu$^4$, Eun-Jung Holden$^1$, Ting Dang$^1$ \\
  $^1$The University of Melbourne \\
  $^2$KAIST \\
  $^3$The University of Auckland \\
  $^4$UNSW Sydney \\
  \texttt{yang.xiao.1@student.unimelb.edu.au}
}

\begin{document}
\maketitle

\begin{abstract}
Large audio language models (LALMs) process both speech and environmental acoustic cues, yet struggle to retain non-speech information across multi-turn interactions. The performance gap between semantic (speech) and acoustic (non-speech) understanding remains poorly understood, and the underlying mechanisms of representation and retrieval are still unclear. This work introduces EnvMem, a controlled multi-turn benchmark designed to study this gap and identify the root causes of failures at the representation (i.e., latent embeddings) and retrieval levels (i.e., attention allocation). We further conduct post-hoc interventions to probe representational structure and attention dynamics. Our results reveal representational trajectory drift as the key failure mode, while showing that attention allocation plays a limited role in explaining the observed degradation. Overall, we provide a systematic framework for analyzing and improving non-linguistic memory in long-context LALMs, shedding light on future data and training design for robust acoustic memory modeling.
\end{abstract}

\section{Introduction}

Large audio language models (LALMs) support end-to-end understanding and generation of spoken conversations \cite{su2025audio, cui2025recent, defossez2024moshi, ding2025kimi}. Early models primarily focused on linguistic content, such as speech transcription and intent recognition \cite{zeng2024glm}. However, spoken interactions also contain rich non-linguistic acoustic context that is absent from transcripts, including environmental sounds such as running water or street traffic. Modeling these non-speech acoustic events is important for grounding dialogue systems in physical environments and improving contextual awareness \cite{sakshi2025mmau, lin2025style}. More recently, LALMs have begun to integrate both speech and non-speech acoustic understanding within unified frameworks, enabling more holistic audio perception and reasoning \cite{kim2026aligning, kumar2026mmau, ma2026mmar,chen2026polybench}.

While acoustic perception has improved, it remains largely constrained to single-turn, short-context settings. Extending this capability to multi-turn conversations is important for applications requiring persistent awareness of the audio scene, such as safety-critical monitoring systems, where maintaining environmental cues (e.g., alarms or hazardous sounds) over time can enable more accurate and reliable decision-making. More fundamentally, it strengthens multimodal autoregressive models’ ability to encode and preserve non-linguistic information over time, supporting the learning of robust and temporally stable audio representations.

Existing work has begun to explore this direction and reports that extending acoustic memory to multi-turn settings is challenging~\cite{gosai2025audio}. Models tend to retain linguistic information from earlier turns but progressively lose environmental acoustic context as dialogue length increases. 
This asymmetric forgetting suggests that semantic content is preserved more reliably than non-speech acoustic context over time. However, prior analyses are often based on entangled speech and acoustic settings, lacking fine-grained quantification of acoustic memory and, in particular, a principled decomposition of the factors contributing to this degradation.
This motivates a closer examination of how environmental acoustic information is retained and processed in multi-turn LALMs, and, more importantly, which factors drive the observed performance gap. 
To address the gaps, in this work, we consider two research questions:
\begin{itemize}[leftmargin=*, itemsep=2pt]
    \item \textbf{RQ1.} What is 
    the performance gap between linguistic and environmental acoustic memory in multi-turn LALMs?
    \item \textbf{RQ2.} Does this acoustic memory failure stem from a representation-level (i.e., information loss during cross-layer propagation) or a retrieval-level (i.e., insufficient attention allocation to early acoustic cues)?
\end{itemize}




To address RQ1, we introduce \textbf{EnvMem}, a controlled multi-turn analytical framework.
Unlike natural datasets where linguistic and acoustic cues are entangled in an unstructured manner, EnvMem systematically couples them by introducing acoustic information in early dialogue turns and probing both semantic and acoustic recall through question answering within the same memory span. This design enables a direct comparison between semantic and acoustic memory under identical conditions.

To address RQ2, we formalize two failure modes, including representation-level degradation and retrieval-level failure, and conduct a white-box diagnostic study to identify the mechanisms underlying acoustic amnesia. 
At the representation level, we use layer-wise linear probing and Centered Kernel Alignment (CKA)~\cite{kornblith2019similarity} to track whether environmental information persists within hidden states or drifts from the retrieval trajectory. 
At the retrieval level, we analyze cross-layer attention weights to determine if the model fails to attend to the initial acoustic segments. 

We further conduct a proof-of-concept intervention study by perturbing latent representations and attention modules to examine how inference-time strategies can improve acoustic memory retention and recovery. This analysis sheds light on efficient post-hoc mechanisms for improving audio representation learning and provides guidance for future training.
In summary, our contributions are:

\begin{itemize}
    \item We construct EnvMem, a multi-turn framework that controls linguistic and acoustic variables, serving as a benchmark for quantifying acoustic memory in long-context LALMs. 
    
    \item We conduct a white-box diagnostic analysis to investigate acoustic amnesia in LALMs, disentangling representation-level drift from retrieval-level attention failures and identifying representational format mismatch as the key bottleneck.
    
\item We provide a proof-of-concept inference-time interventions study
, highlighting the potential of improving representations for enhanced memory retention. 
\end{itemize}

\section{Related Work}

\paragraph{Multi-Turn Evaluation in LALMs.} 
Recent LALMs adopt end-to-end architectures to process speech and acoustic signals~\cite{defossez2024moshi, zeng2024glm, sakshi2025mmau, lin2025style,xiao2026adapting}. However, most models and studies are trained and evaluated in single-turn settings~\cite{yang2024air, wang2025audiobench}. Emerging multi-turn benchmarks such as Audio MultiChallenge~\cite{gosai2025audio} shift toward dialogue-based evaluation but do not explicitly target acoustic memory, instead relying on naturalistic interactions where speech and acoustic cues are entangled across turns. This entanglement may introduce varying acoustic sounds at different dialogue turn positions, scattering acoustic information across the dialogue and removing clear boundaries for analyzing retention and retrieval. As a result, it becomes difficult to disentangle modality-specific memory decay from confounding factors in mixed semantic–acoustic processing. This motivates controlled evaluation frameworks that separate semantic reasoning from acoustic context, enabling isolation of the mechanisms underlying multi-turn acoustic failures.

\begin{figure*}[t]
  \centering
  \includegraphics[width=\textwidth]{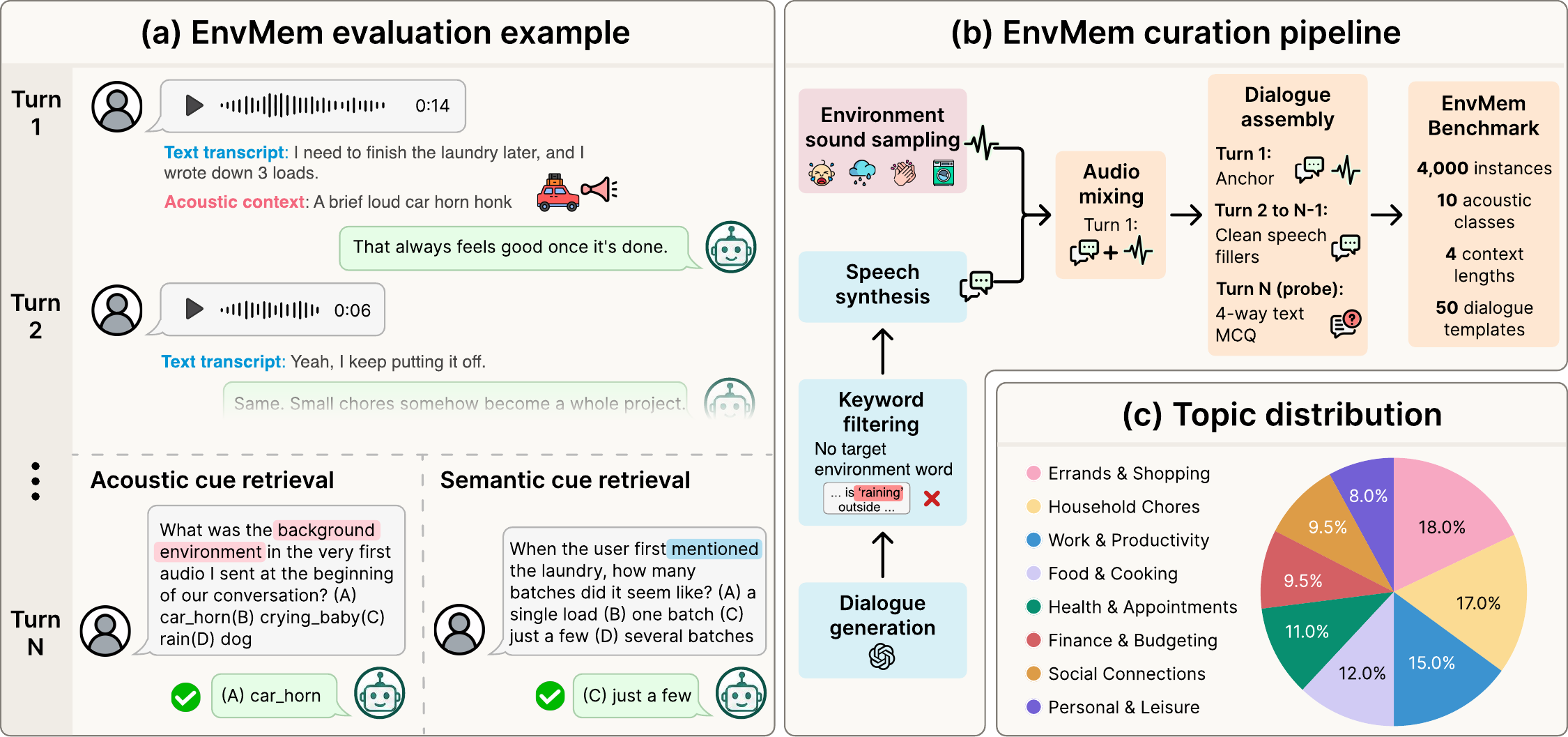}
  \caption{Overview of the EnvMem framework, consisting of three parts. (a) A multi-turn evaluation example illustrating the semantic and acoustic retrieval tasks. (b) The dataset curation pipeline. (c) Dataset topic distribution. }
  \label{fig:chatbox}
\vspace{-10pt}
\end{figure*}

\paragraph{Long-Context Memory in LLMs and LALMs. }
To interpret contextual forgetting in multi-turn models, existing literature often relies on frameworks developed for text-only LLMs. These studies typically attribute long-context degradation either to storage limits within the KV-cache~\cite{mohtashami2023landmark} or to positional decay, such as the ``lost in the middle'' effect~\cite{liu2024lost}. However, LALMs introduce additional acoustic encoders and cross-modal alignment mechanisms, which fundamentally alter how representations are formed and propagated across layers, leaving it unclear whether these failure modes directly transfer to multimodal settings~\cite{yin2026focus}.

Recent empirical evidence shows that non-semantic acoustic cues degrade faster than semantically matched inputs in naturalistic settings~\cite{gosai2025audio}, indicating that acoustic representations follow a distinct cross-layer evolution pattern that generic positional decay models cannot explain.
Instead, acoustic representations appear to undergo modality-specific trajectory drift over extended interactions~\cite{hsu2026anatomy}. It remains unclear whether this drift reflects true information erasure (i.e., irreversible loss of acoustic features) or a retrieval failure (i.e., preserved features that are not routed into the generation pathway).

\section{The EnvMem Analytical Framework}
To investigate multi-turn acoustic memory and explore RQ1, we introduce EnvMem, a controlled analytical framework that disentangles non-semantic acoustic memory from semantic memory.

\subsection{Task Formulation and Design Principles}
Each evaluation sample in EnvMem is a multi-turn spoken dialogue with $N$ turns, denoted as:
\begin{equation}
\small
    \mathcal{D} = \{(u_1, a_2), (u_3, a_4), \dots, (u_{N-1}, a_N),u_N\}
\end{equation} 
where $u_n$ denotes user utterances and $a_n$ represent assistant responses. The dialogue is structured into three phases. As shown in Figure~\ref{fig:chatbox}(a), the first user utterance $u_1$ serves as an acoustic anchor: a speech utterance mixed with one environmental sound (e.g., rain) at a fixed signal-to-noise ratio of 10 dB, embedding the target non-semantic cue exclusively in this initial turn. The intermediate user utterances $u_k$ ($1 < k < N$) act as fillers that extend the context without introducing new environmental acoustics. The final utterance $u_N$ is a text-based probe question asking the model to recall either the environmental sound or a semantic fact from $u_1$. 
For the baseline context ($N=2$) used to analyze the single-turn result, the model receives the anchor audio $u_1$, followed immediately by the probe $u_2$.

This design follows two principles to resolve the entanglement of prior benchmarks. First, restricting the acoustic cue to $u_1$ establishes a clear temporal anchor, ensuring that performance degradation at $u_N$ reflects cross-turn memory rather than repeated exposure of acoustic events. 
Second, each sequence supports both acoustic and semantic probes over the same audio context, allowing direct comparison under identical inputs. 

\begin{figure*}[t!]
  \centering
  \includegraphics[width=\textwidth]{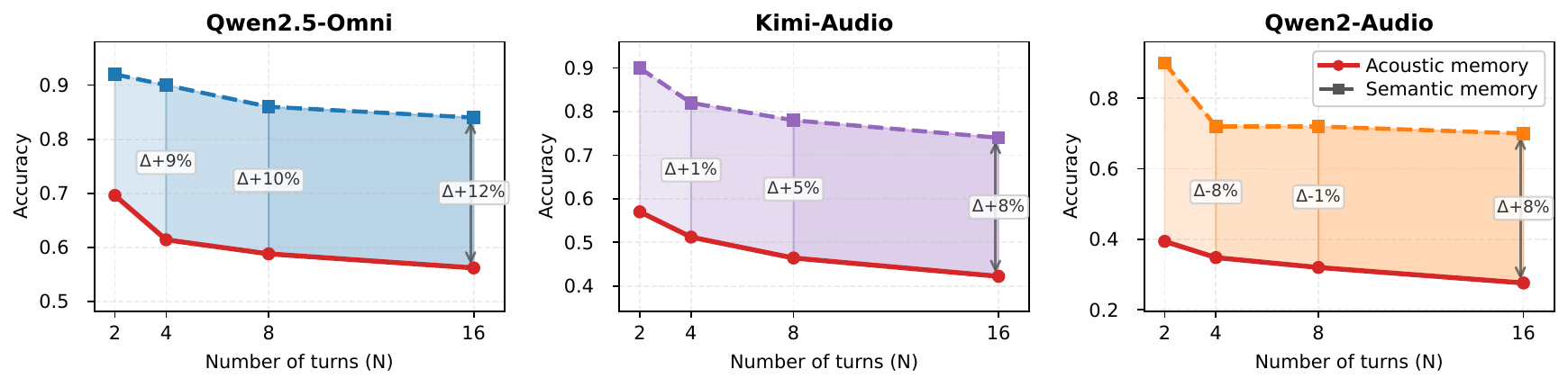}
  \vspace{-6mm}
  \caption{Acoustic vs. semantic memory performance over increasing dialogue turns ($N$). The consistent modality gap ($\Delta$) indicates a shared deficiency in retaining non-linguistic environmental cues in extended LALM contexts.}
  \label{fig:acoustic-semantic}
  \vspace{-5mm}
\end{figure*}




\subsection{Dataset Construction and Acoustic Events}
The acoustic events are sampled from 10 classes of the ESC-50 dataset \cite{piczak2015esc}. These classes cover four coarse acoustic families: weather/water, machine, vocal-bio, and impulse. This mix provides both continuous background textures and transient sound events. The class list and per-class counts are detailed in Appendix~\ref{app:dataset}.

The conversational text is constructed using multi-turn dialogue templates generated by GPT-4o across eight everyday topics, as shown in Figure~\ref{fig:chatbox}(c). To support a paired evaluation design, GPT-4o generates an acoustic probe that queries the background sound and a semantic probe that queries a factual entity mentioned in $u_1$. Both probes are formulated as four-way multiple-choice questions. Keyword filtering is further applied to remove templates whose intermediate utterances contain words related to the target environment, preventing leakage of environmental cues. 

To prevent evaluation confounds related to speaker or paralinguistic variations, as shown in Figure 1(b), all utterances (including $u_1$ and fillers) are synthesized using a single TTS voice (Kokoro \texttt{af\_heart}~\cite{kokoro-tts} and resampled to 16 kHz to match the LALM input rate.
The benchmark covers the 10 acoustic classes across 4 context lengths with 50 independent dialogue templates per condition. This results in 2,000 test cases for each of the acoustic and semantic probes, yielding a total of 4,000 evaluation instances. 

\subsection{Experimental Setup}

To establish a baseline for EnvMem, we evaluate three LALMs: 
Qwen2.5-Omni~\cite{Qwen2.5-Omni}, Qwen2-Audio~\cite{chu2024qwen2audiotechnicalreport}, and the Kimi-Audio~\cite{ding2025kimi}. These models represent recent open-weight architectures that support native multimodal fusion without relying on cascaded ASR components.

For each test instance, the model receives an $N$-turn audio sequence with a corresponding 4-way multiple-choice question for either acoustic or semantic querying, as shown in Figure 1(a). The system prompt restricts outputs to the selected option or its letter (e.g., “Option A”). Accuracy is used as the primary metric, computed by extracting the predicted choice from the model response and comparing it to the ground truth. To disentangle cross-turn memory decay from initial perception differences, we define a relative degradation metric, $\Delta(N)$, measuring the difference between the accuracy drop in acoustic (a) and semantic (s) tasks relative to their single-turn ($N=2$) baselines: 

{\small
\begin{equation}
\begin{split}
\Delta(N) = \frac{\text{acc}_a(2) - \text{acc}_a(N)}{\text{acc}_a(2)}
          - \frac{\text{acc}_s(2) - \text{acc}_s(N)}{\text{acc}_s(2)}
\end{split}
\end{equation}
}
$\Delta > 0$ indicates that acoustic memory degrades faster than semantic memory relative to the model's baseline capabilities.

\subsection{Performance Analysis}

To quantify acoustic memory loss (RQ1), we evaluate the selected LALMs across increasing conversational context lengths ($N \in \{2, 4, 8, 16\}$). 

\paragraph{Perception Gap. }
As shown in Figure \ref{fig:acoustic-semantic}, we observe a modality gap between spoken semantic retention and environmental sound understanding across models. The $N=2$ baseline reveals that part of this gap comes from initial perception differences: Qwen2-Audio shows strong imbalance (0.39 acoustic vs. 0.90 semantic accuracy), indicating weaker encoding of non-linguistic cues rather than purely long-context failure. In contrast, Qwen2.5-Omni and Kimi-Audio achieve higher acoustic baselines (0.70 and 0.57), enabling more reliable analysis of cross-turn degradation.



\paragraph{Multi-Turn Propagation. }
As $N$ increases from 2 to 16, acoustic memory degrades more rapidly than semantic memory despite identical temporal spans. For Qwen2.5-Omni, semantic accuracy drops from 0.92 to 0.84, while acoustic accuracy declines from 0.70 to 0.56, yielding $\Delta = +12\%$ at $N=16$. Kimi-Audio shows a similar trend with $\Delta = +8\%$.


Qwen2-Audio shows a negative $\Delta$ at $N=4$ due to weak single-turn performance, but the absolute acoustic–semantic gap grows with context and converges to $\Delta = +8\%$ at $N=16$. Overall, positive $\Delta$ at long contexts indicates modality-specific degradation, where environmental acoustic cues are more susceptible to cross-turn memory loss than spoken semantics.

\begin{figure*}[t!]
  \centering
  \begin{subfigure}[b]{0.48\textwidth}
    \centering
    \includegraphics[width=\textwidth]{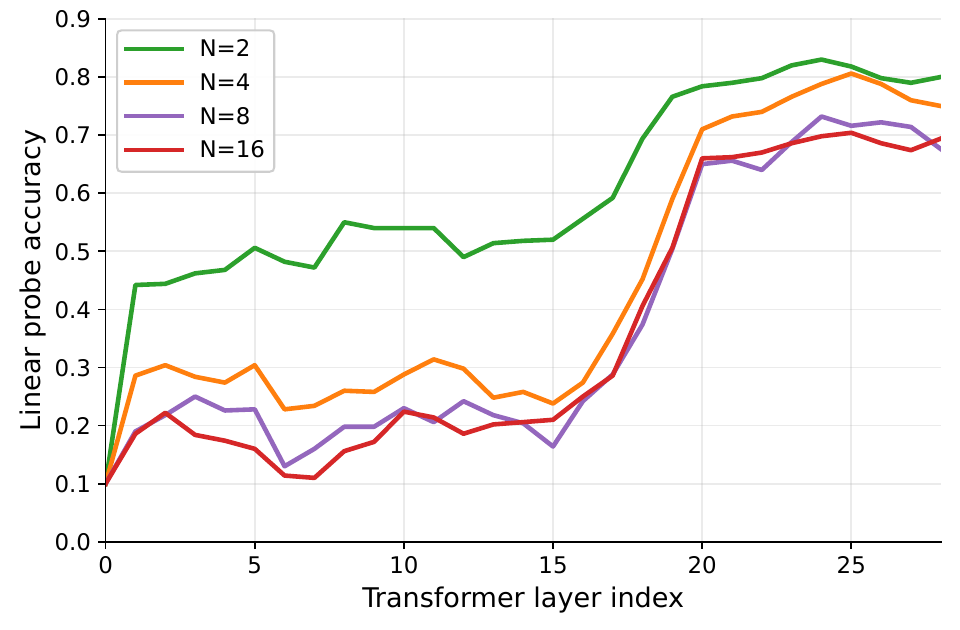}
    \vspace{-5mm}
    \caption{All samples.}
    \label{fig:probe_all}
  \end{subfigure}
  \hfill
  \begin{subfigure}[b]{0.48\textwidth}
    \centering
    \includegraphics[width=\textwidth]{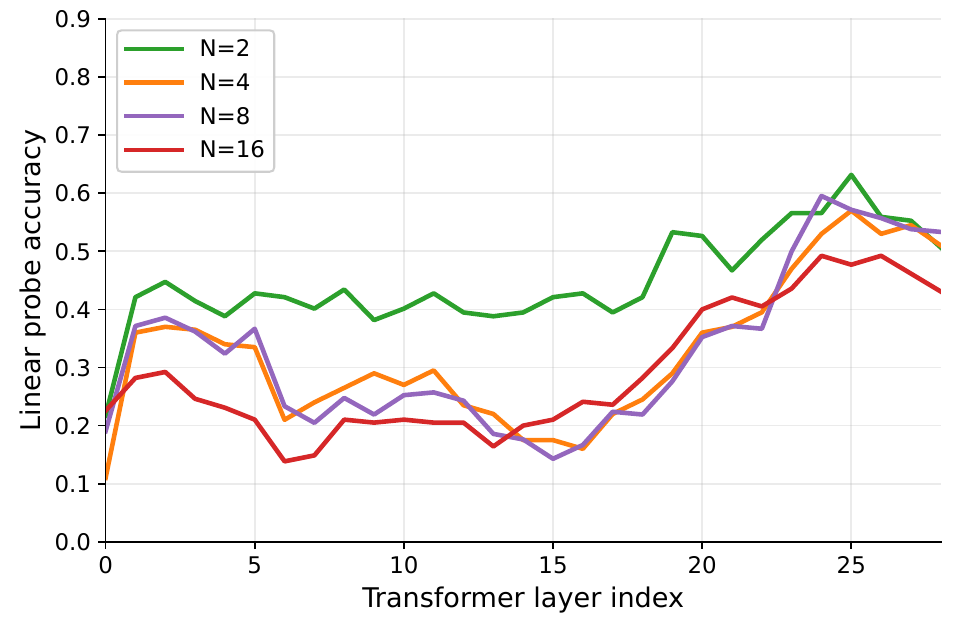}
    \vspace{-5mm}
    \caption{Failed samples only.}
    \label{fig:probe_failed}
  \end{subfigure}
  \vspace{-3mm}
  \caption{Layer-wise linear probe accuracy by context length $N$. (a) Full evaluation set. (b) Restricted to trials where the model produced an incorrect prediction. Probe accuracy well above the
  10-way chance level ($0.10$) in deep layers under both conditions confirms that acoustic information remains decodable within hidden states.
  }
  \vspace{-5mm}
  \label{fig:probe}
\end{figure*}

\section{Diagnosing Acoustic Memory Decay}

\label{sec:diagnosis}

To address RQ2, we formalize the failure mechanisms of acoustic memory across two dimensions: (i) the representation level (\S\ref{sec:4-1}), determining whether acoustic information is preserved within the model's hidden states as they propagate across layers, and (ii) the retrieval level (\S\ref{sec:4-2}), assessing whether attention mechanisms correctly route that information to the output.

\subsection{Representation-Level Analysis}
\label{sec:4-1}
\subsubsection{Linear Probing}
We employ a layer-wise linear probing protocol for 10-class environmental sound classification: for each layer $\ell \in \{0, \dots, 27\}$, we train a linear classifier on the hidden state of the final query token. Higher probing accuracy indicates stronger preservation of acoustic information in the representation. The final query token is the model's active reasoning state during generation, so probing it measures whether distant acoustic anchors can be retrieved and integrated into the generation pathway at a given depth, rather than the static storage of localized acoustic tokens.
As illustrated in Figure 3, the linear decodability of the target environmental class 
reveals three mechanistic phenomena.

\paragraph{Deep-Layer Preservation.} 
Across all dialogue lengths $N \in {2, 4, 8, 16}$, acoustic information is weakly represented in early and middle layers but becomes recoverable in upper transformer blocks ($\ell \ge 20$). At $N=16$, probe accuracy reaches ~70\% by layer 25, compared to ~80\% for the $N=2$ baseline, indicating only a modest degradation with longer context.
Critically, this decodability persists even when conditioned on the failed subset. As shown in Figure~\ref{fig:probe_failed}, when restricting evaluation to incorrect outputs, layer-25 probe accuracy remains ~48\% under $N=16$, well above chance (10\%). 
Therefore, acoustic information remains preserved in hidden states but is only reliably accessible in deep layers, suggesting that representational capacity is not the primary limitation underlying acoustic memory loss. 

\paragraph{Delayed Acoustic Integration.} Beyond eventual recovery in deep layers, context length shifts where acoustic information is integrated in the network. Under short context ($N=2$), acoustic cues are incorporated early, with query tokens reaching ~0.50 accuracy by layer 5 and gradually improving thereafter. In contrast, for longer contexts ($N=8$ and $N=16$), mid-layer performance remains low (0.20 up to layer 15), followed by a sharp increase in layers 16–21. This indicates that long-range acoustic retrieval is deferred to deeper layers as context length increases. Whether this shift is driven by attention allocation is analyzed in §4.2.

\begin{figure}[t!]
  \centering
  \includegraphics[width=\columnwidth]{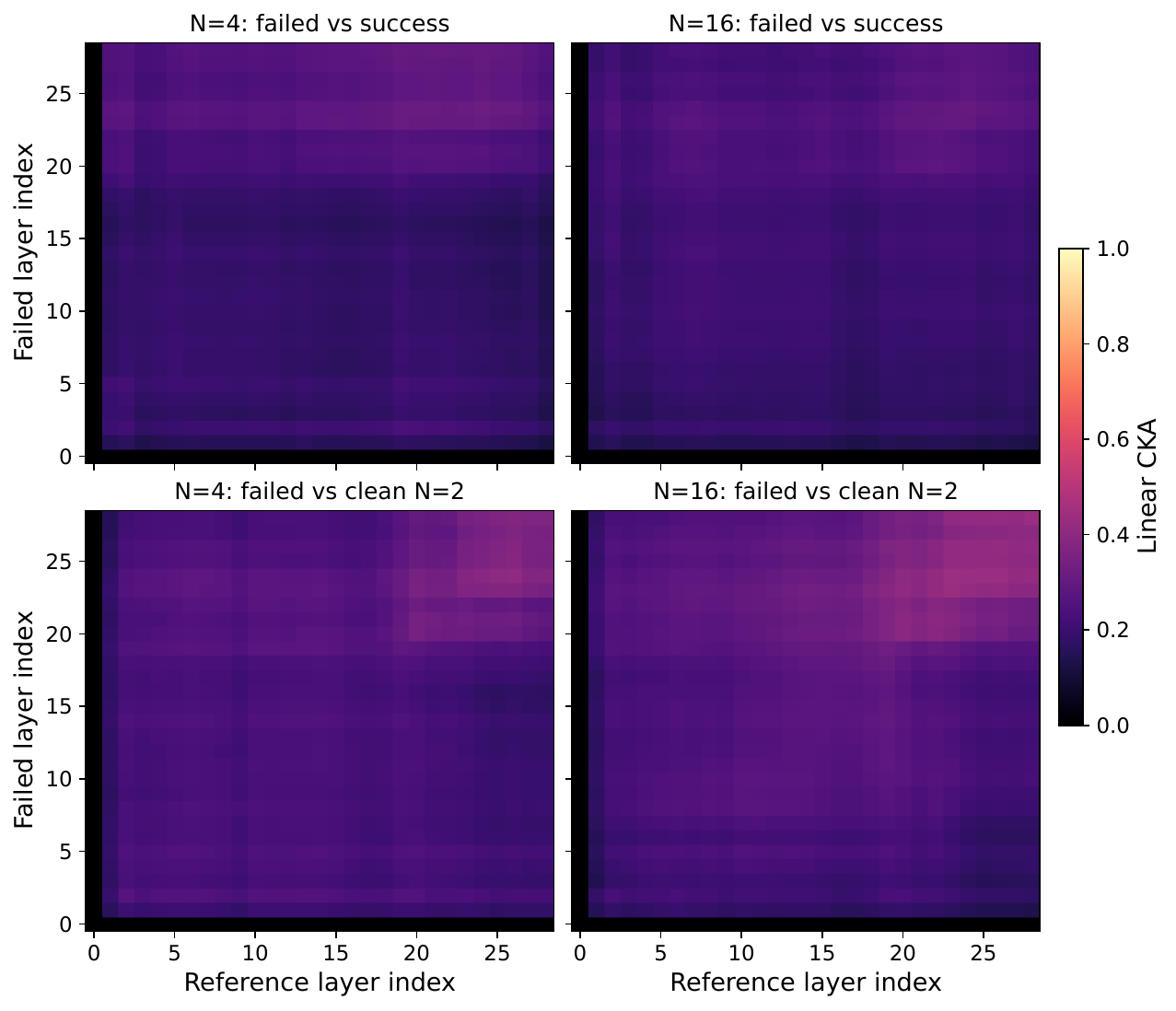}
  \vspace{-8mm}
\caption{Cross-layer CKA alignment heatmaps between failed, successful,
and short ($N=2$) trial hidden states at the decoding position.
Each cell reports the linear CKA score between layer $\ell_a$ (y-axis)
and layer $\ell_b$ (x-axis). Top row: failed vs.\ same-$N$ success.
Bottom row: failed vs.\ short $N=2$. Results for $N=8$ are consistent with $N=16$ and are moved to the Appendix~\ref{app:cka}.}
\vspace{-4mm}
\label{fig:cka_heatmap}
\end{figure}

\subsubsection{Cross-Layer Trajectory (CKA)}
\label{sec:4-1-2}

We further use Centered Kernel Alignment (CKA)~\cite{kornblith2019similarity} to evaluate how these hidden representations evolve across successive layers. For each context length, we compare CKA across three conditions, each matched on the same environmental sound class
: failed long-context trials, successful long-context trials, and a short-context baseline ($N=2$), which we refer to as short. Comparing failed against successful trials shows how a failed representation differs from a correct one, while comparing both against the short baseline reveals whether a failed representation instead reverts to the short-context regime. 

The heatmaps (Figure~\ref{fig:cka_heatmap}) reveal three phases. In Phase~I (layers~0--5), CKA values remain near zero across all three conditions, indicating that the representations have not yet differentiated by trial outcome or context length. In Phase~II (layers~6--20), failed and successful trajectories diverge. We quantify this with the paired difference, defined as:
\begin{equation}
\footnotesize
    \delta = \mathrm{CKA}(\text{failed}_N, \text{short}_{2}) - \mathrm{CKA}(\text{failed}_N, \text{success}_N)
\end{equation} 
averaged over the block. The positive $\delta$ across all settings ($N=4$: 0.054, $N=8$: 0.055, $N=16$: 0.067) indicates that failed long-context representations are consistently closer to the short-context (short) baseline than to successful long-context trajectories. This suggests that failures are not simply corrupted long-context processing, but instead \emph{partially collapse toward a short-context computation regime}. In Phase~III (layers~20--28), a high-similarity block emerges but peaks off the main diagonal, indicating that successful long-context integration still occurs, but with a delayed or phase-shifted alignment relative to the short trajectory.

\paragraph{Observational evidence for representational trajectory drift.} The probing and CKA results collectively demonstrate that acoustic information is preserved rather than erased. Even on failed trials, a linear probe recovers the target class well above chance in deep layers. The failure instead stems from the structural format of this representation. Specifically, on failed trials, the hidden states revert to a short-memory pathway in Phase~II and only aggregate the acoustic signal in deep layers, causing a phase shift relative to a successful retrieval. We term this as \emph{representational trajectory drift}. This suggests that the main bottleneck is a mismatch between the structure of the learned representation and the expectations of the decoding pathway, rather than information loss.

\begin{figure}[t!]
  \centering
  \includegraphics[width=0.8\columnwidth]{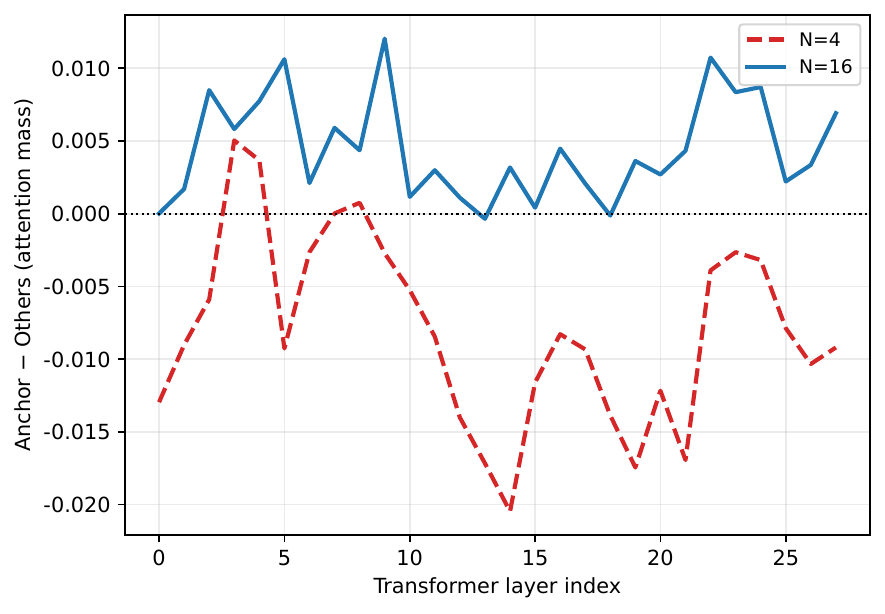}
  \caption{Anchor attention gap by layer (Qwen2.5-Omni). Negative values mean the model attends less to the anchor than to fillers.}
  \label{fig:attention_gap}
\end{figure}

\subsection{Retrieval-Level Analysis}
\label{sec:4-2}


Section~\ref{sec:4-1} indicated that anchor acoustic information is recoverable from deep hidden states, yet the late-layer representation deviates from the successful trajectory. A natural hypothesis is that this deviation is caused by attention: in extended contexts, attention may be diverted away from anchor-audio tokens, so that recoverable information never reaches the output. We test this hypothesis with three attention diagnostics and find that attention allocation does not explain the failure.

\paragraph{Attention gap  on the anchor turn. }
For each sample, we compute the difference between i) the average attention mass assigned 
to the anchor-turn token span and ii) the mean attention mass assigned to other user-turn spans (fillers) at every transformer layer $\ell$. Positive values indicate higher attention to the anchor turn; negative values indicate lower attention relative to filler turns. Figure~\ref{fig:attention_gap} 
on Qwen2.5-Omni shows that 
at $N=4$, the anchor gap is predominantly negative across the network (reaching a minimum around layers 14--15), indicating less attention to the anchor than to filler turns. In contrast, at $N=16$, the gap becomes mostly positive ($+0.005$ to $+0.010$), showing stronger attention to the anchor despite longer context. The gap does not decrease with increasing $N$ and does not track accuracy, suggesting that attention allocation alone does not explain memory decay.

\paragraph{Attention concentration. }
To test whether failures are due to more diffuse attention, we compute $\mathrm{cov}_{90}$, the number of turns (or tokens) needed to capture 90\% of attention at the decoding position, where lower values indicate more concentrated attention. For each $N$, we pair failed and successful samples from the same class and compare $\Delta_d = \mathrm{cov}{90}^{\text{failed}} - \mathrm{cov}{90}^{\text{success}}$. As shown in Table~\ref{tab:cov90}, $\Delta_d$ is close to zero and inconsistent across $N$ (turn level: $+0.04$, $-0.10$, $-0.10$), and negligible at the token level. Overall, failed trials are not more attention-diffuse than successful ones, suggesting attention concentration does not explain the performance gap.


\begin{table}[t!]
\centering
\small
\setlength{\tabcolsep}{6pt} 
\caption{Attention concentration on matched failed vs success samples, measured by 
$\mathrm{cov}_{90}$. 
}
\label{tab:cov90}
\begin{tabular}{lcccc}
\toprule
\textbf{Metric} & \textbf{$N$ }& \textbf{ Failed} & \textbf{Success} & \textbf{$\Delta_d$} \\
\midrule
\multirow{3}{*}{\centering$\mathrm{cov}_{90}$ (turns)}  & 4  & 2.71  & 2.67  & $+0.04$ \\
                             & 8  & 6.15  & 6.25  & $-0.10$ \\
                             & 16 & 10.15 & 10.25 & $-0.10$ \\ \cmidrule{2-5}
\multirow{3}{*}{\centering$\mathrm{cov}_{90}$ (tokens)} & 4  & 64.8  & 62.5  & $+2.4$ \\
                             & 8  & 96.6  & 94.8  & $+1.8$ \\
                             & 16 & 136.5 & 130.6 & $+5.9$ \\
\bottomrule

\end{tabular}
\\[2pt]
\end{table}


\paragraph{Attention Allocation Statistics Do Not Predict Failure. }
These findings confirm that macro-level results (Figure~\ref{fig:attention_gap}) show no uniform dilution of target attention with longer contexts, while micro-level statistics (Table~\ref{tab:cov90}) indicate that attention concentration alone does not explain success or failure.



\section{Causal Validation}
\label{sec:causal}
We design post-hoc interventions at both the representation and attention levels. If failures stem from corrupted representations, directly restoring the representation should recover correct predictions, whereas manipulating attention should have limited effect, and vice versa.

\begin{figure*}[t]
  \centering
  \includegraphics[width=0.9\textwidth]{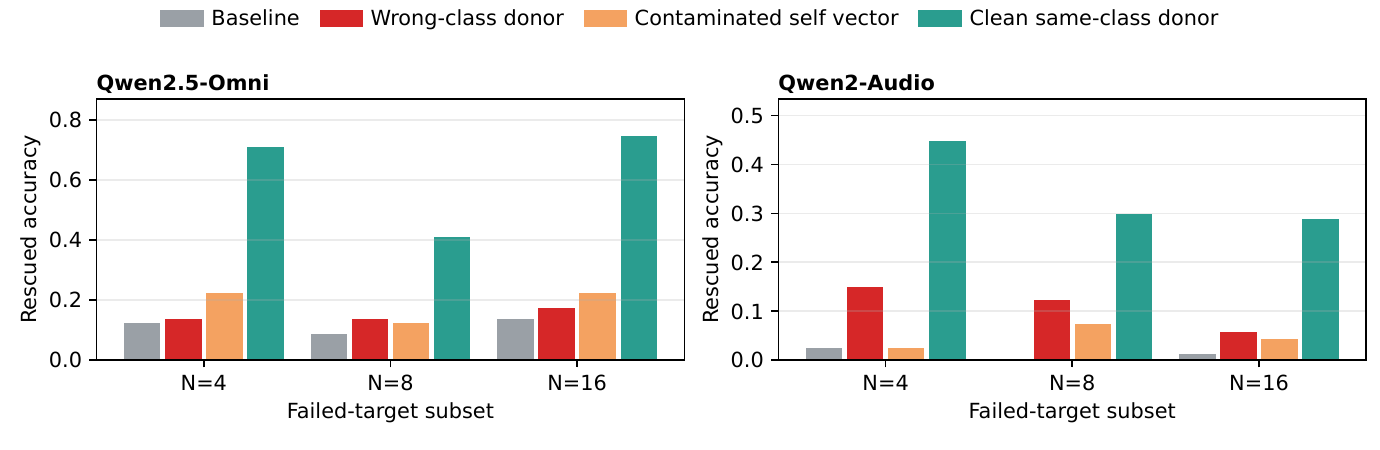}
  \vspace{-6mm}
  \caption{Activation patching on failed targets. Only clean same-class donors rescue failed predictions; wrong-class donors and contaminated self-vectors remain near baseline, ruling out a generic-perturbation effect.}
  \label{fig:activation-patching}
  \vspace{-5mm}
\end{figure*}

\subsection{Activation Patching at Deep Layers} 
\label{sec:5-2}

\paragraph{Setup. }

To evaluate the representation hypothesis, we apply activation patching by overwriting the hidden state at the decoding position at layer $\ell_{\mathrm{patch}}$ with a donor vector, while keeping all other states unchanged and continuing the forward pass. We test three donor types on failed trials: (i) a wrong-class donor from a different environmental class at the same $N$, (ii) a contaminated same-class donor from another failed trial, and (iii) a clean same-class donor from a correctly answered long-context trial. Following probing results, $\ell_{\mathrm{patch}}$ is set to the best-performing layer (layer 25 for Qwen2.5-Omni), with additional layer-offset sweeps to assess depth sensitivity (Appendix~\ref{app:patching-details}).

\paragraph{Donor specificity.}
As shown in Figure~\ref{fig:activation-patching}, donor identity critically determines whether the failed prediction is recovered. For Qwen2.5-Omni at $N=16$, clean-donor patching boosts accuracy from a near-chance baseline ($0.13$) to $0.75$, whereas the wrong-class donor ($0.17$) and the contaminated self-vector ($0.22$) remain near baseline. The same pattern holds at $N \in \{4, 8\}$. Qwen2-Audio shows the identical pattern, with clean-donor accuracy of $0.30$--$0.45$ across $N$ while both control donors stay near baseline. The rescue is therefore neither a generic perturbation effect (the wrong-class donor fails) nor a by-product of injecting any same-class state (the model's own failed representation is ineffective). Only a clean, format-compatible representation restores correct decoding.

\paragraph{Takeaway. }
Maintaining a format-compatible representation is the decisive factor for successful long-context acoustic retrieval. Because overwriting the deep-layer representation alone recovers the majority of failed predictions, the failure is attributable to the representation level and does not reflect irreversible information loss.

\subsection{Attention Manipulation Does Not Recover Failed Predictions}
\label{sec:5-1}

\begin{table}[t]
\centering
\normalsize
\caption{Attention-mask interventions on the failed subset (Qwen2.5-Omni, 80 samples per $N$). Baseline rows report accuracy (\%); intervention rows report $\Delta$ vs.\ baseline with paired bootstrap 95\% CIs.}
\label{tab:attn_intervention}
\resizebox{\linewidth}{!}
{\footnotesize
\begin{tabular}{@{}lccc@{}}
\toprule
& $N\!=\!4$ & $N\!=\!8$ & $N\!=\!16$ \\
\midrule
Baseline (\%) & $12.5${\scriptsize$~[6.2,20.0]$} & $12.5${\scriptsize$~[6.2,20.0]$} & $15.0${\scriptsize$~[7.5,22.5]$} \\
\midrule
\multicolumn{4}{c}{Intervention $\Delta$ (\%)} \\
\midrule
Anchor amp. & $+1.3${\scriptsize$~[-5.0,+7.5]$} & $+1.3${\scriptsize$~[-5.0,+7.5]$} & $+1.2${\scriptsize$~[-6.2,+8.8]$} \\
Filler supp. & $0.0${\scriptsize$~[-6.2,+7.5]$} & $+2.5${\scriptsize$~[-5.0,+11.2]$} & $0.0${\scriptsize$~[-7.5,+7.5]$} \\
Random ctrl. & $-1.3${\scriptsize$~[-7.5,+5.0]$} & $0.0${\scriptsize$~[-6.2,+7.5]$} & $+1.2${\scriptsize$~[-6.2,+8.8]$} \\
\bottomrule
\end{tabular}}
\end{table}

\paragraph{Setup}
To test the routing hypothesis on failed cases, we modify the last-layer additive attention mask at the decoding query row under three settings: (i) anchor amplification, increasing attention to the anchor-turn span; (ii) filler suppression, removing attention to all non-anchor user turns; and (iii) a magnitude-matched random-span boost as a control. Details are provided in Appendix~\ref{app:attn-sweep}.



\paragraph{Result.}
Table~\ref{tab:attn_intervention} reports per-condition accuracy and $\Delta$ relative to the baseline, with paired-bootstrap 95\% confidence intervals (CIs). All interventions yield $\Delta$ values whose CIs include zero and overlap with the random-span control. Filler suppression shows small, inconsistent effects ($+2.5$ at $N=8$, $0$ at $N \in {4,16}$). A sweep over layer offsets ${-8,-4,0,+4}$ and boost magnitudes ${2,4,8}\times$ yields a maximum $\Delta$ of $+3.7$, comparable to the $+1.3$ from the random-span control (Appendix~\ref{app:attn-sweep}).

\paragraph{Takeaway.}
Consistent with the attention analysis in Section~\ref{sec:4-2}, modifying attention mass yields minimal performance changes, indicating that misrouted attention is unlikely the primary cause of failure.

\paragraph{Summary.}
The outcomes of the two interventions are asymmetric. Restoring the deep-layer representation recovers the majority of failed predictions, whereas manipulating attention does not. This causally confirms that multi-turn acoustic amnesia is driven by representational trajectory drift rather than retrieval-level misrouting, and that the failure is corrigible by acting on the representation itself.


\section{Discussion}
\label{sec:6}

We translate the analytical findings and intervention results into actionable guidance for evaluating and mitigating multi-turn acoustic amnesia.

\paragraph{Implications for Benchmark Design.} EnvMem provides an initial step toward constructing controlled benchmarks that disentangle acoustic and semantic memory. By holding conversation scripts, voice, and turn structure constant while varying only the probe target, it enables isolated analysis of modality-specific memory effects and reduces confounding between content recall and acoustic recall. This design offers practical guidelines for building datasets aimed at controlled diagnosis of multimodal models, particularly in long-context audio settings. It supports more reliable evaluation of where failures arise within the processing pipeline. Future research could expand this framework to broader acoustic conditions, more diverse speaker settings, and more complex conversational structures.


\paragraph{Diagnosing the Bottleneck.} The findings highlight a representational-level bottleneck, indicating that current LALMs remain limited in preserving acoustic information relative to semantic memory. This gap is likely driven by pretraining and post-training pipelines that underemphasize acoustic events and insufficiently couple semantic and acoustic signals in real-world data. Beyond diagnosis, this study suggests directions for future training design with a stronger emphasis on acoustic event modeling. Formulating these memory effects at both representation and attention levels enables more systematic evaluation and comparison of multimodal models, providing a unified framework for failure analysis, and generalization to other multimodal settings.


\paragraph{Directions for Mitigation.} While our post-hoc intervention strategy provides insights into how representations and attention can be adjusted to improve performance, it also motivates future work on enhancing representation learning directly, beyond post-hoc structural corrections. In particular, more effective strategies are needed to better integrate acoustic representations during training to strengthen acoustic memory in future models.

\section{Conclusion}
In this work, we systematically investigated the mechanisms underlying acoustic memory in long-context audio-language models. Through representation-level analysis, attention inspection, and corresponding targeted interventions, we show that acoustic information is largely preserved in latent representations but becomes functionally misaligned with the decoding process in long contexts. Our results suggest that the primary bottleneck is not information loss but rather a representational and routing mismatch that emerges in deeper layers. These findings provide both diagnostic tools and mechanistic insights for understanding multi-turn acoustic memory, and highlight the need for training strategies that more tightly couple acoustic and semantic representations in long-context settings.

\clearpage
\section*{Limitations}
First, EnvMem is built on controlled, synthetic multi-turn dialogues: utterances are TTS-synthesized with a single voice, and each dialogue contains a single environmental anchor at a fixed signal-to-noise ratio. This design is intentional, as it isolates cross-turn acoustic retrieval from confounds such as speaker variation and overlapping acoustic events. However, it does not capture the full variability of natural conversational audio, including multiple or time-varying acoustic events, spontaneous speech, and channel noise. Extending the framework to more naturalistic conditions is a natural next step. Second, our interventions are inference-time, post-hoc analyses designed to causally diagnose the bottleneck rather than to serve as deployable mitigation methods. Translating these diagnostic insights into training-time objectives that durably strengthen acoustic memory remains an open problem.

\section*{Ethics Statement}
The benchmark uses synthetic dialogue generation, TTS, and public environmental sounds for diagnostic model evaluation.
Potential misuse includes cross-turn profiling of users' physical environments (e.g., inferring home/work context, routine, or sensitive background conditions) from seemingly harmless conversational audio.
Such capabilities can increase surveillance and privacy risks in remote-assistant settings if deployed without clear consent and retention limits.
Potential misuse also includes anthropomorphic over-interpretation of memory failures.
We release this benchmark as an evaluation and safety analysis tool, not as a deployment recipe for user profiling.


\bibliography{custom}

\clearpage
\appendix
\section{Dataset Details}
\label{app:dataset}

\begin{table}[h!]
\centering
\small
\resizebox{\columnwidth}{!}{
\begin{tabular}{cccc}
\toprule
\textbf{Sound Family} & \textbf{ESC-50 Class} & \textbf{Per Cell ($N$)} & \textbf{Total Samples} \\
\midrule
\multirow{3}{*}{Weather / Water} & rain & 50 & 200 \\
 & thunderstorm & 50 & 200 \\
 & sea\_waves & 50 & 200 \\
\midrule
\multirow{3}{*}{Machine} & engine & 50 & 200 \\
 & vacuum\_cleaner & 50 & 200 \\
 & car\_horn & 50 & 200 \\
\midrule
\multirow{2}{*}{Vocal-Bio} & dog & 50 & 200 \\
 & crying\_baby & 50 & 200 \\
\midrule
\multirow{2}{*}{Impulse} & clock\_tick & 50 & 200 \\
 & crackling\_fire & 50 & 200 \\
\midrule
Total & 10 Classes & 50 & 2,000 \\
\bottomrule
\end{tabular}}
\caption{Detailed distribution of acoustic event classes and sound families in EnvMem.}
\label{tab:acoustic_classes}
\end{table}

This appendix provides the specific names of the ten acoustic classes, the complete keyword filtering lexicon, and concrete multi-turn dialogue examples utilized in the EnvMem framework.

\subsection{Acoustic Class List and Counts}
Table~\ref{tab:acoustic_classes} details the ten specific classes selected from the ESC-50 dataset mapped to their respective sound families. Each class contains exactly 50 base templates per context length ($N$), resulting in a completely balanced dataset.

\subsection{Keyword Filtering Lexicon}
To prevent the textual leakage of acoustic cues, intermediate filler turns (Turns 1 to $N-1$) are filtered using a strict blacklist of 42 control words. This lexicon is divided into four categories. During pipeline curation, text from intermediate turns is lowercased and tokenized via a regular expression matching alphabetic strings. Any template yielding a non-empty intersection with this lexicon is discarded. Out of 208 initial candidate templates generated by GPT-4o, 38 were filtered out, ensuring strict semantic isolation.

\subsection{System Prompt}

To simulate real-world daily interactions and prevent representational overfitting, we utilized GPT-4o to generate conversational scripts. The exact system prompt used to instruct the LLM for generating the multi-turn base templates and their semantic parallels is provided below.  

\begin{tcolorbox}[
    title={Dialogue Generation System Prompt},
    halign title=center,
    colback=white, 
    colframe=gray!80, 
    boxrule=0.8pt, 
    arc=3mm, 
    fonttitle=\bfseries, 
    fontupper=\small, 
    left=1.5ex, right=1.5ex, top=1ex, bottom=1ex 
]

You are a dialogue writer for a speech-memory benchmark. Produce natural English multi-turn conversations as valid JSON only. Do not include markdown fences, commentary, or any text outside the JSON array.

\vspace{0.5em}
\noindent\textbf{Task:} Generate \texttt{\{batch\_size\}} dialogue samples as a JSON array. \\
Each sample must be an object with fields \texttt{"id"}, \texttt{"turns"}, and \texttt{"semantic\_parallel"}. \\
Both \texttt{"turns"} and \texttt{"semantic\_parallel"} contain exactly 16 turns with this schema:
\begin{center}
    \texttt{\{"turn\_id": \textit{int}, "role": "user"|"assistant", "text": \textit{string}\}}
\end{center}

\noindent\textbf{Instructions:}
\begin{enumerate}[leftmargin=*, itemsep=2pt, parsep=0pt, topsep=2pt]
    \item Roles must alternate strictly: odd turns are \texttt{"user"}, even turns are \texttt{"assistant"}.
    \item Turn 1 in \texttt{"turns"} must describe an ordinary daily task or concern. It must not imply location, setting, environment, sound, or scene.
    \item Turns 2--15 should be natural casual conversation that drifts away from turn 1 without abrupt topic jumps.
    \item Turn 16 in both variants must be exactly \texttt{<PROBE\_PLACEHOLDER>} with role \texttt{"user"}.
    \item \texttt{"semantic\_parallel"} should closely mirror the style and trajectory of \texttt{"turns"}, but turn 1 must embed one concrete text-only fact suitable for later probing (e.g., a name, number, weekday, date, room number, or item count).
    \item Do not include any environment or scene cues anywhere in turns 1--15 of either variant.
    \item Keep wording conversational, concise, and varied across samples; avoid repeating the same dialogue structure.
\end{enumerate}

\vspace{0.5em}
\noindent\textbf{Forbidden words in turns 1--15 of either variant:} \\
\texttt{rain, sunny, weather, outside, inside, noisy, quiet, loud, echo, background, cafe, restaurant, office, home, street, traffic, dog, cat, baby, music, engine, car, wind, thunder, fire, wave, sea, storm, animal}

\vspace{1em}
Respond with a JSON array only --- nothing else.
\end{tcolorbox}

\section{Linear Probing Details}
\label{app:pro}

\subsection{Feature Extraction}
We train an independent linear classifier for every combination of conversational memory load $N \in \{2, 4, 8, 16\}$ and hidden-state layer index $\ell$. The input features are the final query-token activations, extracted at the probe turn immediately prior to generation, from a single frozen forward pass.

\subsection{Probe Architecture and Training}
Each probe is instantiated as a standard multinomial logistic classifier (a single linear layer with bias, parameterized as $\mathbb{R}^{d \times 10}$). The probes are trained to predict the ten environmental sound classes (such as car\_horn, clock\_tick, and rain). Optimization is performed using the AdamW optimizer with a learning rate of $3 \times 10^{-3}$ and a weight decay of $1 \times 10^{-4}$. Models are trained with a cross-entropy loss for 120 epochs on a single GPU.

\subsection{Evaluation Protocol}
Probe performance is estimated using stratified $k$-fold cross-validation, where $k = \min(5, N_{\min})$ and $N_{\min}$ represents the minimum sample count across all classes. Folds with fewer than two examples per class are excluded. Splits are generated with a fixed random seed (random\_state=0), and we report the mean fold accuracy against a theoretical chance level of 10\%.

\subsection{Data Scale and Evaluation Subsets}
We conduct our linear probing analysis using the complete, unified hidden-state bank of 2,000 valid trials. This comprises 500 trials per $N$, which is perfectly balanced to exactly 50 samples per class per $N$. Random-label control probes evaluated on this set consistently cluster near the theoretical chance level of 10\%.

\begin{figure*}[t!]
  \centering
  \begin{subfigure}[b]{0.48\textwidth}
    \centering
    \includegraphics[width=\textwidth]{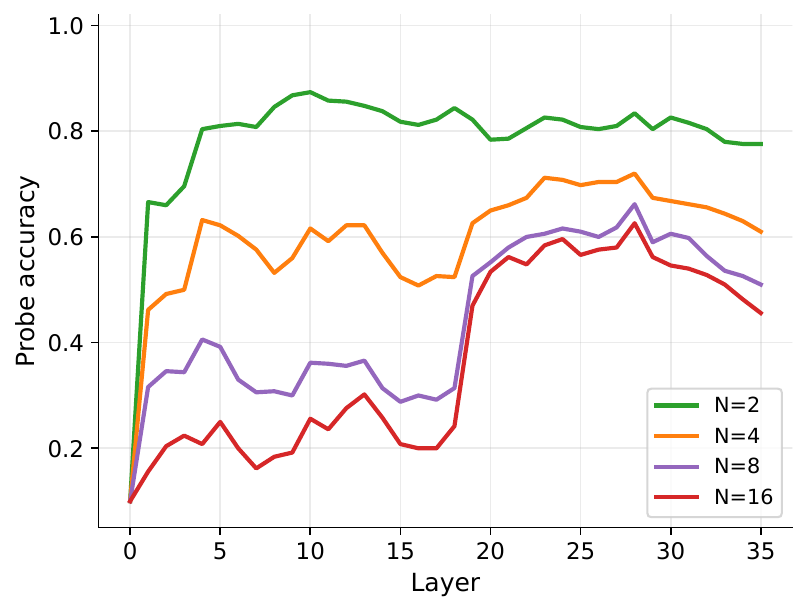}
    \vspace{-5mm}
    \caption{Kimi-Audio}
    \label{fig:probe_kimi}
  \end{subfigure}
  \hfill
  \begin{subfigure}[b]{0.48\textwidth}
    \centering
    \includegraphics[width=\textwidth]{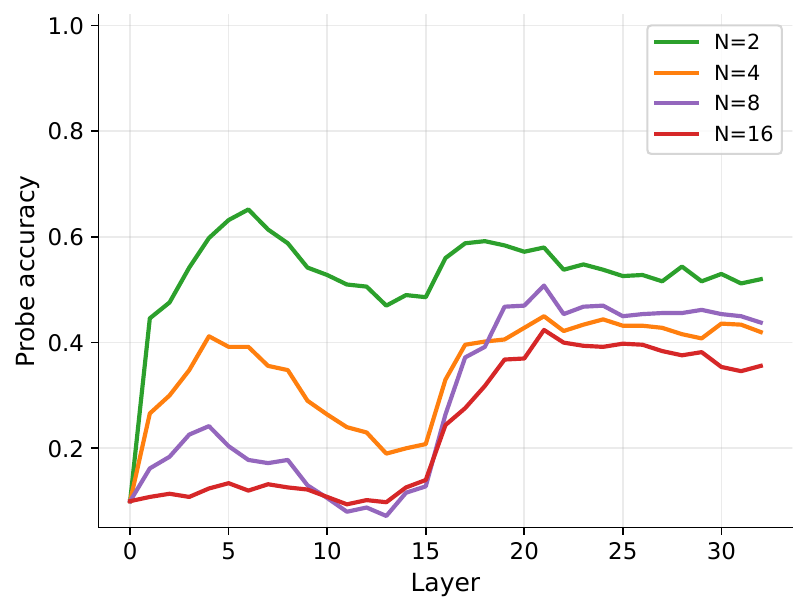}
    \vspace{-5mm}
    \caption{Qwen2-Audio}
    \label{fig:probe_qwen}
  \end{subfigure}
  \vspace{-3mm}
  \caption{Cross-model comparison of layer-wise linear probe accuracy by context length $N$ on the full evaluation set. Consistent with the primary findings, both (a) Kimi-Audio and (b) Qwen2-Audio demonstrate robust deep-layer preservation of acoustic cues, alongside a pronounced delay in acoustic integration as context length increases. This confirms that representational trajectory drift is a shared behavior across different LALM architectures.}
  \vspace{-5mm}
  \label{fig:probe_cross_model}
\end{figure*}

\subsection{Cross-Model Generalization}

While Section 4.1.1 establishes the primary representation-level dynamics using Qwen2.5-Omni, we extend our linear probing analysis to Kimi-Audio and Qwen2-Audio to verify if these memory failures are architecture-agnostic. As illustrated in Figure~\ref{fig:probe_cross_model}, the layer-wise probing trajectories of both Kimi-Audio and Qwen2-Audio closely mirror the phenomena observed in the primary model. Specifically, both models exhibit the same delayed acoustic integration under extended contexts ($N \ge 8$), where acoustic information remains largely undecodable in early-to-middle layers before experiencing a sharp recovery in the deepest transformer blocks. This cross-model consistency confirms that deep-layer preservation and representational trajectory drift are fundamental, shared architectural bottlenecks in current LALMs, rather than artifacts of any single training paradigm.

\section{CKA Analysis Details}
\label{app:cka}

\subsection{Linear CKA}
We compute \textbf{linear CKA}~\cite{kornblith2019similarity} between sets of
hidden-state vectors. Given two matrices $X, Y \in \mathbb{R}^{n \times d}$,
let $\tilde{X} = X - \bar{X}$ denote column-wise mean-centred features; the
similarity is
\begin{equation}
  \text{CKA}(X,Y)=\frac{\lVert\tilde{X}\tilde{X}^{\!\top}\odot\tilde{Y}\tilde{Y}^{\!\top}\rVert_{F}}
                        {\lVert\tilde{X}\tilde{X}^{\!\top}\rVert_{F}\,
                         \lVert\tilde{Y}\tilde{Y}^{\!\top}\rVert_{F}}.
\end{equation}
Linear CKA is invariant to orthogonal transformation and isotropic scaling,
making it suitable for comparing representations across layers and trials.

\paragraph{Representations.}
We extract the last-token hidden state at each transformer layer (indices
0--28, $d{=}3584$ for the Qwen2.5-Omni, using the probe turn as input.
The number of layers is model-specific; the same procedure is applied to
Qwen2-Audio and Kimi-Audio.

\paragraph{Pair construction.}
For each $N \in \{4,8,16\}$ we form two sets of class-matched pairs
(up to 80 pairs per set, random seed 2026):
\begin{itemize}
  \item \emph{failed vs.\ same-$N$ success}: a failed sample paired with a
        correct sample from the same acoustic class and the same $N$;
  \item \emph{failed vs.\ clean $N{=}2$}: a failed sample paired with a
        correct sample from the same acoustic class at $N{=}2$.
\end{itemize}

\paragraph{Cross-layer matrix.}
We compute a cross-layer matrix whose $(i,j)$ entry is
$\text{CKA}(H^{(i)}_{\text{failed}},\, H^{(j)}_{\text{ref}})$,
where $H^{(l)}$ collects the layer-$l$ last-token vectors across all pairs.
The diagonal ($i{=}j$) gives layer-wise representational similarity.
The \emph{estimated drop layer} is defined as
$\hat{l}_{\mathrm{drop}}=\arg\min_{l}\,(\mathrm{diag}[l{+}1]-\mathrm{diag}[l])$,
the layer at which diagonal similarity falls most sharply. It is a
visualization aid only: because layer~0 is fixed near zero and the diagonal
rises steeply in the first few layers, $\hat{l}_{\mathrm{drop}}$ typically
falls in the early stack and does \emph{not} localize the trajectory drift,
which is instead characterized by the Phase~II and Phase~III analysis in
Section~\ref{sec:4-1-2}.

\subsection{Per-Model Results}
Figures~\ref{fig:cka_scanline_qwen25omni} report
the CKA scanline alignment for the three evaluated LALMs. The main text analyzes Qwen2.5-Omni at $N\in\{4,16\}$; here we
additionally provide the $N{=}8$ condition and the corresponding results for
Qwen2-Audio and Kimi-Audio. Across all three models, the three-phase pattern
holds: failed and successful trajectories separate in the mid-stack
(Phase~II), and a phase-shifted high-similarity block emerges in the deep
layers (Phase~III). This indicates that representational trajectory drift is a
shared property of multi-turn acoustic memory rather than an artifact of a
single architecture.

\begin{figure}[t!]
  \centering
  \includegraphics[width=\columnwidth]{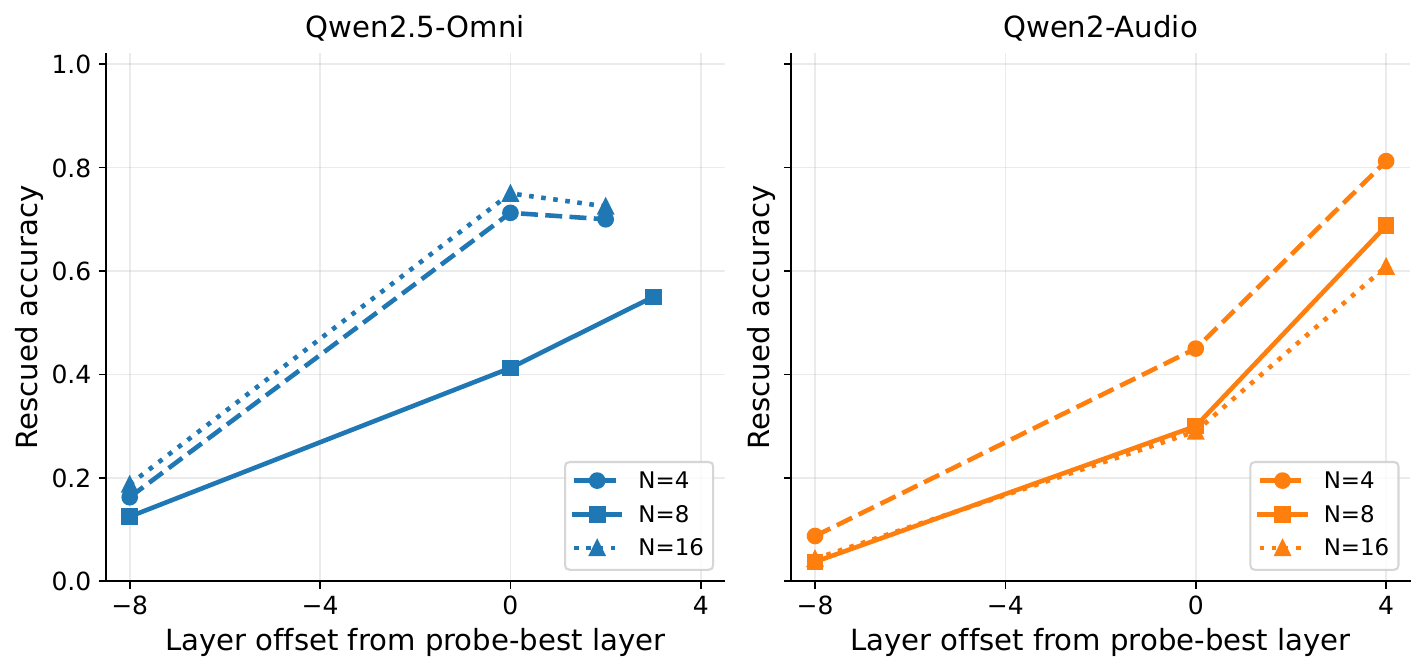}
  \caption{Activation-patch rescue accuracy across layer offsets relative to
  the probe-best layer, for Qwen2.5-Omni and Qwen2-Audio.}
  \label{fig:patch_sweep}
\end{figure}

\begin{figure*}[t!]
  \centering
  \includegraphics[width=0.8\linewidth]{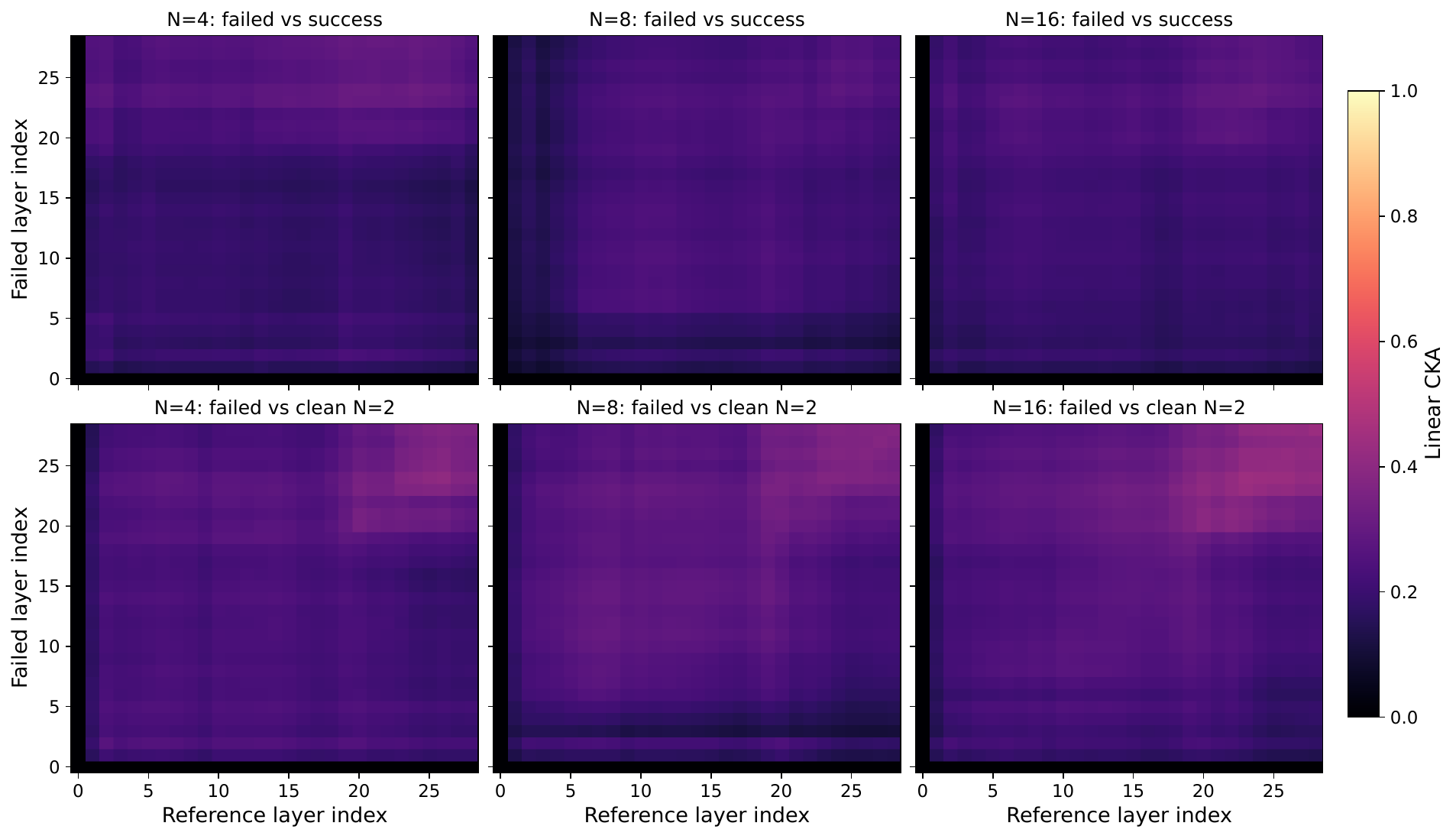}
  \vspace{-2mm}
  \caption{CKA scanline alignment for Qwen2.5-Omni, including the $N{=}8$
  condition deferred from the main text.}
  \label{fig:cka_scanline_qwen25omni}
\end{figure*}

\begin{figure*}[t!]
  \centering
  \includegraphics[width=0.8\linewidth]{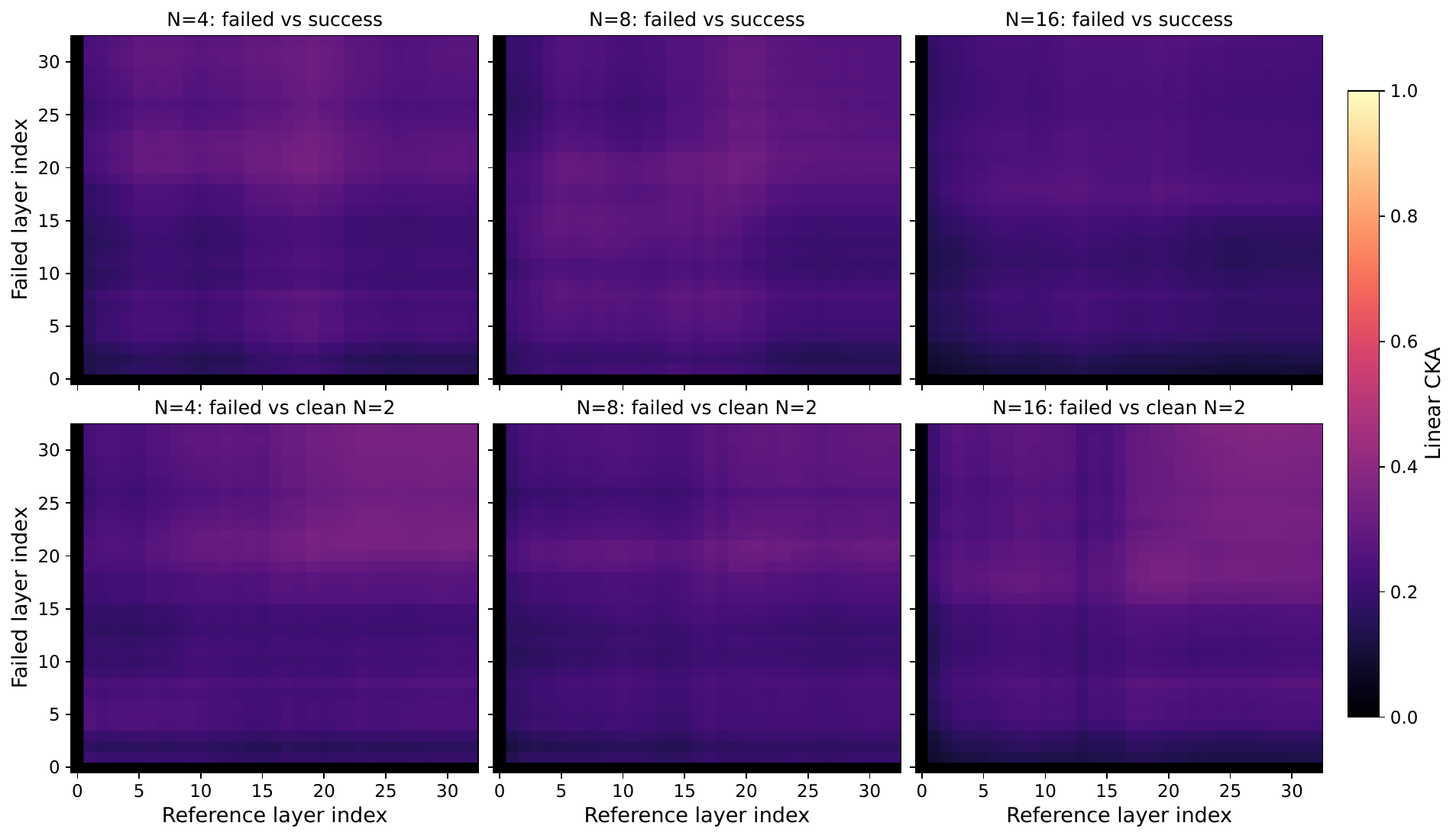}
  \vspace{-2mm}
  \caption{CKA scanline alignment for Qwen2-Audio. The three-phase pattern is
  consistent with Qwen2.5-Omni.}
  \label{fig:cka_scanline_qwen2audio}
\end{figure*}


\section{Activation Patching Details}
\label{app:patching-details}

\paragraph{Patch target.}
For each failed sample (target), we select a class-matched donor from the
\emph{same-$N$ success} pool and from the \emph{clean $N{=}2$ success} pool. The donor's hidden state at the
probe-best layer $\ell_{\mathrm{probe}}$ is injected into the target's forward
pass by replacing the last-token activation at a patch layer
$\ell_{\mathrm{patch}}$ via a PyTorch forward hook:
\begin{equation}
  h^{(\ell_{\mathrm{patch}})}_{-1} \;\leftarrow\; h^{(\ell_{\mathrm{probe}})}_{\mathrm{donor}},
\end{equation}
where $-1$ denotes the last (probe) token position, and
$h^{(\ell_{\mathrm{probe}})}_{\mathrm{donor}}$ is taken from the donor's own
forward pass. All other positions and layers are left untouched.
$\ell_{\mathrm{probe}}$ is the layer with the highest linear-probe accuracy for
each $N$. All confidence intervals are
nonparametric percentile bootstrap 2{,}000 resamples.

\paragraph{Layer offset sweep.}
To assess sensitivity to the exact injection depth, we sweep
$\ell_{\mathrm{patch}} \in \{\ell_{\mathrm{probe}}-8,\; \ell_{\mathrm{probe}},\;
\ell_{\mathrm{probe}}+\delta\}$, with $\delta = \min(4,\, L-1-\ell_{\mathrm{probe}})$
capping the positive offset at the final layer. The sweep assumes
$\ell_{\mathrm{probe}}+\delta$ stays within the same transformer stack; this
holds for the two Qwen models but not for Kimi-Audio, whose 28-layer main
stack is routed through a separate MiMo branch before the text head. We
therefore report the single-point result at $\ell_{\mathrm{probe}}$ for all
three models and restrict the sweep to Qwen2.5-Omni and Qwen2-Audio.
For both Qwen models, rescue is strong throughout the mid-to-upper band
(offset $0$ to $+4$): the best offset is architecture-dependent (offset $0$
for Qwen2.5-Omni, offset $+4$ for Qwen2-Audio), but accuracy stays high across
this range. Patching far below the probe-best layer (offset $-8$) yields
negligible improvement ($\Delta \le {+}7$\,pp) for both models, confirming
that recoverable acoustic information is concentrated in the mid-to-upper
layers rather than in early processing.

\section{Attention Manipulation Details}
\label{app:attn-sweep}

\paragraph{Intervention mechanism.}
All interventions are applied via a forward pre-hook on the self-attention module of the target transformer layer. The hook modifies the four-dimensional attention-mask tensor (batch by heads by sequence by sequence) exclusively at the last query position, leaving all other rows unchanged. Three intervention types are implemented.

The first intervention type is anchor amplification. This applies a multiplier $v \in \{2, 4, 8\}$ by adding $\log v$ to the pre-softmax attention logits over the anchor-turn key span $[s, e)$:
\[
    m_{-1,\,s:e} \mathrel{+}= \log v
\]

The second intervention type is filler suppression. This applies a scaled penalty to all non-anchor key spans, determined by a parameter $v \in \{0.5, 0.25, 0\}$. For all $k \neq 0$:
\[
    m_{-1,\,s_k:e_k} \mathrel{+}=
    \begin{cases} \log v & v > 0 \\ -10^{9} & v = 0 \end{cases}
\]

The third intervention type is a random-span control. This applies anchor amplification with $v=4$ to a randomly chosen non-anchor turn span, serving as a position-matched null control.

\paragraph{Sweep and statistics.}
Each intervention is applied at layer offsets $\ell \in \{\ell_{\mathrm{probe}}-8, \ell_{\mathrm{probe}}-4, \ell_{\mathrm{probe}}, \ell_{\mathrm{probe}}+4\}$, using exactly 80 failed targets per context length $N$ evaluated under a fixed random seed. To test whether the change in accuracy ($\Delta\mathrm{acc}$) is greater than zero, we employ a one-sided paired percentile bootstrap utilizing 2,000 resamples. Per-trial binary outcomes are resampled with replacement, the difference $\Delta$ is recomputed, and the one-sided $p$-value is defined as the fraction of resamples yielding $\Delta \leq 0$.


\end{document}